\documentclass[prd,preprint,superscriptaddress,showpacs,byrevtex]{revtex4}
\usepackage{bm}
\def\fsl#1{\setbox0=\hbox{$#1$}           
   \dimen0=\wd0                                 
   \setbox1=\hbox{/} \dimen1=\wd1               
   \ifdim\dimen0>\dimen1                        
      \rlap{\hbox to \dimen0{\hfil/\hfil}}      
      #1                                        
   \else                                        
      \rlap{\hbox to \dimen1{\hfil$#1$\hfil}}   
      /                                         
   \fi}                                         %
\newcommand{\be}{\begin{equation}}
\newcommand{\ee}{\end{equation}}
\newcommand{\bea}{\begin{eqnarray}}
\newcommand{\eea}{\end{eqnarray}}
\newcommand{\beq}{\begin{equation}}
\newcommand{\eeq}{\end{equation}}
\newcommand{\beqs}{\begin{eqnarray}}
\newcommand{\eeqs}{\end{eqnarray}}

\begin{document}
\title{ Gauge Invariant Noether's Theorem and The Proton Spin Crisis }
\author{Gouranga C Nayak }\thanks{G. C. Nayak was affiliated with C. N. Yang Institute for Theoretical Physics in 2004-2007.}
\affiliation{ C. N. Yang Institute for Theoretical Physics, Stony Brook University, Stony Brook NY, 11794-3840 USA}
\date{\today}
\begin{abstract}
Due to proton spin crisis it is necessary to understand the gauge invariant definition of the spin and orbital angular momentum of the quark and gluon from first principle. In this paper we derive the gauge invariant Noether's theorem by using combined Lorentz transformation plus local gauge transformation. We find that the notion of the gauge invariant definition of the spin (or orbital) angular momentum of the electromagnetic field does not exist in Dirac-Maxwell theory although the notion of the gauge invariant definition of the spin (or orbital) angular momentum of the electron exists. We find that the gauge invariant definition of the spin angular momentum of the electromagnetic field in the literature is not correct because of the non-vanishing surface term in Dirac-Maxwell theory although the corresponding surface term vanishes for linear momentum. We also show that the Belinfante-Rosenfeld tensor is not required to obtain symmetric and gauge invariant energy-momentum tensor of the electron and the electromagnetic field in Dirac-Maxwell theory.
\end{abstract}
\maketitle
\pagestyle{plain}

\pagenumbering{arabic}

\section{Introduction}

The spin of the proton at rest is $\frac{1}{2}$. When the proton is in motion, like that at high energy colliders, its helicity (which is the projection of the proton spin along its direction of motion) is conserved with the quantized value $\pm \frac{1}{2}$. In the naive parton model it was predicted that the proton spin is carried by the quarks (plus antiquarks) inside the proton \cite{jaffe,jaffe1}. However, the famous European muon collaboration (EMC) experiment at CERN \cite{emc} revealed that the total contribution to the proton spin from the quarks (plus antiquarks) is almost zero. This is known as the "proton spin crisis" which is one of the most important unsolved problem in particle physics.

After the EMC experiment, other experiments have also confirmed similar results \cite{allspin}, including the recent RHIC results at ${\sqrt s}_{NN}$ = 510 GeV polarized proton-proton collisons \cite{rhic,rhic1}. RHIC experiment also involves heavy-ion collisions to study quark-gluon plasma \cite{n1,n2}. In future, the electron-ion collider (EIC) \cite{eic} is expected to provide the precise insight to the spin structure of the proton. At present, the total quarks (plus antiquarks) spin contribution $S_q$ to the proton spin is $\sim \frac{1}{6}$ \cite{totals} and the total gluons helicity contribution $S_g$ to proton spin is $\sim \frac{1}{10}$ \cite{totals} and the rest of the proton spin may be from the orbital angular momentum $L_q$ of the quarks (plus antiquarks) and from the orbital angular momentum $L_g$ of the gluons.

Assuming total angular momentum conservation in physics one usually writes
\bea
\frac{1}{2} = S_q + S_g + L_q + L_g
\label{aspn}
\eea
where in terms of experimentally measured polarized quark (plus antiquark) distribution function $\Delta q(x)$ one has
\bea
S_q=\frac{1}{2}~\int_0^1 dx ~\Delta q(x)
\label{bspn}
\eea
and in terms of experimentally measured polarized gluon distribution function $\Delta g(x)$ one has
\bea
S_g=\frac{1}{2}~\int_0^1 dx ~\Delta g(x)
\label{cspn}
\eea
where $x$ is the longitudinal momentum fraction of the parton with respect to proton.

Since the RHIC experiment \cite{rhic,rhic1} measure spin polarized gluon distribution function $\Delta g(x)$ inside the proton one expects that the gauge invariant spin distribution function of the gluon is measured at the experiments. However, it is well known in the gauge theory that only the spin angular momentum $S_q$ of the quark is gauge invariant but the spin angular momentum $S_g$ of the gluon is not gauge invariant. Take for example the Maxwell theory where the spin angular momentum of the electromagnetic field obtained from the Noether's theorem is given by [see eq. (\ref{ecnem}) for the derivation]
\bea
{\vec S}_\gamma = \int d^3x ~{\vec E}(x)\times {\vec A}(x)
\label{dspn}
\eea
which is not gauge invariant where ${\vec E}(x)$ is the electric field which is gauge invariant and ${\vec A}(x)$ is the electromagnetic vector potential which is not  gauge invariant.

The gauge invariant definition of the spin dependent gluon distribution function proposed in \cite{jaffe1,manh} corresponds to eq. (\ref{dspn}) in the light-cone gauge $A^+=0$. However, in any other gauge it does not correspond to gluon spin angular momentum that is obtained from Noether's theorem. Hence we do not have a gauge invariant definition of spin dependent gluon distribution function $\Delta g(x)$ in QCD at high energy colliders from the first principle. This is because the gauge invariant definition of the spin dependent gluon distribution function must give the same value in any gauge.

In addition to this there have been various (gauge invariant and gauge non-invariant) definitions of the spin angular momentum $S_g$ of the gluon, orbital angular momentum $L_g$ of the gluon and orbital angular momentum $L_q$ of the quark in the literature \cite{jaffe1,ji1,wata,hata,gold}.

Hence it is necessary to understand the gauge invariant definition of the spin and orbital angular momentum of the quark and gluon from first principle.

The first principle method to understand the conservation of angular momentum in physics is via Noether's theorem using Lorentz transformation. For example the spin angular momentum $S_\gamma$ of the electromagnetic field in eq. (\ref{dspn}) is obtained from the Noether's theorem by using Lorentz transformation [see eq. (\ref{ecnem})]. The Noether's theorem in Maxwell theory is given by
\bea
\partial_\mu [F^{\mu \nu}(x) \delta A_\nu(x)] - \delta x^\mu \partial_\mu {\cal L}(x)=0
\label{s5n1}
\eea
where $\delta A_\mu(x)$ is the functional differential of $A_\mu(x)$ and
\bea
{\cal L}(x) =-\frac{1}{4}F_{\mu \nu}(x)F^{\mu \nu}(x),~~~~~~~~~~~~~~~~F_{\mu \nu}(x) =\partial_\mu A_\nu(x)-\partial_\nu A_\mu(x).
\label{p5n}
\eea
However under the Lorentz transformation the functional differential $\delta A_\mu(x)$ is not gauge invariant. For example, under the(infinitesimal) Lorentz transformation the functional differential $\delta A_\mu(x)$ is given by
\bea
\delta A_\mu(x) = -F_{\nu \mu}(x)~ \delta x^\nu -\partial_\mu [A_\nu(x) ~\delta x^\nu]
\label{fspn}
\eea
which is not gauge invariant where the general coordinate transformation is given by
\bea
x'_\mu=x_\mu + \epsilon_{\mu \nu}x^\nu +\Delta_\mu = x_\mu +\delta x_\mu.
\label{gspn}
\eea
Note that every thing in the Noether's theorem in eq. (\ref{s5n1}) is gauge invariant except the functional differential $\delta A_\mu(x)$ as given by eq. (\ref{fspn}). Hence when the gauge non-invariant functional differential $\delta A_\mu(x)$ from eq. (\ref{fspn}) is used in eq. (\ref{s5n1}) we obtain the gauge non-invariant expression of the energy-momentum tensor and the gauge non-invariant expression of the angular momentum tensor from the Noether's theorem. This is the origin of the appearance of the gauge non-invariance in Noether's theorem in Maxwell theory which gives the gauge non-invariant spin angular momentum and the gauge non-invariant orbital angular momentum of the electromagnetic field from the Noether's theorem. This implies that the gauge non-invariant conserved quantities in Noether's theorem arises in Maxwell theory because the Lorentz transformation of $A_\mu(x)$ is considered but the local gauge transformation of $A_\mu(x)$ is not considered while deriving the Noether's theorem.

Hence in order to predict the gauge invariant conserved quantities in gauge theory from the first principle it is necessary to implement the local gauge transformation along with the Lorentz transformation to derive the Noether's theorem.

In this paper we derive the gauge invariant Noether's theorem in gauge theory by using the combined Lorentz transformation plus local gauge transformation [see section \ref{gi}]. We find that the notion of the gauge invariant definition of the spin (or orbital) angular momentum of the electromagnetic field does not exist in Dirac-Maxwell theory although the notion of the gauge invariant definition of the spin (or orbital) angular momentum of the electron exists. We find that the gauge invariant definition of the spin angular momentum of the electromagnetic field in the literature \cite{jaffe1,ji1,wata,hata,gold} is not correct because of the non-vanishing surface term [see eq. (\ref{pvbt})] in Dirac-Maxwell theory although the corresponding surface term vanishes for linear momentum. We also show that the Belinfante-Rosenfeld tensor is not required to obtain symmetric and gauge invariant energy-momentum tensor of the electron and the electromagnetic field in Dirac-Maxwell theory. We find that although the high energy collider experiments have measured the spin dependent gluon distribution function inside proton but we do not have a gauge invariant definition of the spin dependent gluon distribution function in QCD consistent with the gauge invariant Noether's theorem.

The paper is organized as follows. In section II we briefly review the gauge non-invariant definition of angular momentum from Noether's theorem in Dirac-Maxwell theory. In section III we discuss non-vanishing boundary term and gauge non-invariant definition of spin and orbital angular momentum in Dirac-Maxwell theory. In section IV (V) we derive the gauge invariant Noether's theorem in Maxwell (Dirac) theory using combined Lorentz transformation plus local gauge transformation. Section VI contains the derivation of the gauge invariant Noether's theorem in Dirac-Maxwell Theory by using combined Lorentz transformation plus local gauge transformation. In section VII we derive the symmetric and gauge invariant definition of the energy-momentum tensor of the electron plus electromagnetic field in Dirac-Maxwell theory without requiring the Belinfante-Rosenfeld tensor. In section VIII we derive the gauge invariant definition of angular momentum from gauge invariant Noether's theorem in Dirac-Maxwell theory using combined Lorentz transformation plus local gauge transformation. Section IX contains conclusions.

\section{ Gauge Non-Invariant Angular Momentum From Noether's Theorem in Dirac-Maxwell Theory }

In this section we will briefly review the gauge non-invariant definition of the angular momentum obtained from the Noether's theorem by using Lorentz transformation.
\subsection{ Gauge Non-Invariant Orbital Angular Momentum of Electromagnetic Field in Maxwell Theory From Noether's Theorem}

Under translation [without rotation, $\epsilon_{\mu \nu}=0$, see eq. (\ref{gspn})] by using eq. (\ref{fspn}) in (\ref{s5n1}) in Maxwell theory we find the continuity equation
\bea
\partial_\mu T^{\mu \nu}(x)=0
\label{gnem}
\eea
where the gauge non-invariant energy-momentum tensor of the electromagnetic field in Maxwell theory obtained from Noether's theorem using Lorentz transformation is given by
\bea
&&T^{\mu \nu}(x)= F^{\mu \lambda }(x) F_\lambda^{~~\nu}(x) +\frac{1}{4}g^{\mu \nu}  F_{\lambda \delta}(x)F^{\lambda \delta}(x) -F^{\mu \lambda }(x)  \partial_\lambda A^\nu(x).
\label{hnem}
\eea

Under rotation [without translation, $\Delta_\mu =0$, see eq. (\ref{gspn})] by using eq. (\ref{fspn}) in (\ref{s5n1}) in Maxwell theory we find the continuity equation
\bea
\partial_\mu J^{\mu \nu \lambda}(x)=0
\label{anem}
\eea
where the gauge non-invariant third rank tensor $J^{\mu \nu \lambda}(x)$ is given by
\bea
J^{\mu \nu \lambda}(x)=-F^{\mu \lambda }(x)A^\nu(x)+F^{\mu \nu }(x)A^\lambda(x)- T^{\mu \nu}(x) x^\lambda+ T^{\mu \lambda}(x) x^\nu
\label{bnem}
\eea
which gives the gauge  non-invariant angular momentum tensor of the electromagnetic field
\bea
M^{\mu \nu}(x)=J^{0 \mu \nu}(x)=-F^{0 \nu }(x)A^\mu(x)+F^{0 \mu }(x)A^\nu(x)- T^{0 \mu}(x) x^\nu + T^{0 \nu}(x) x^\mu
\label{cnem}
\eea
where the gauge non-invariant invariant $T^{\mu \nu}(x)$ is given by eq. (\ref{hnem}). We write eq. (\ref{cnem}) as
\bea
M^{\mu \nu}(x)=S^{\mu \nu }(x)+L^{\mu \nu }(x)
\label{acnem}
\eea
where the spin angular momentum tensor of the electromagnetic field is given by
\bea
S^{\mu \nu}(x)=-F^{0 \nu}(x)A^\mu(x)+F^{0 \mu }(x)A^\nu(x)
\label{bcnem}
\eea
and the orbital angular momentum tensor of the electromagnetic field is given by
\bea
L^{\mu \nu}(x)=- T^{0 \mu}(x) x^\nu + T^{0 \nu}(x) x^\mu
\label{ccnem}
\eea
where the gauge non-invariant invariant $T^{\mu \nu}(x)$ is given by eq. (\ref{hnem}).

The spin angular momentum ${\vec S}$, the orbital angular momentum ${\vec L}$ and the total angular momentum ${\vec J}$ are given by
\bea
S^i=\frac{1}{2}\int d^3x~\epsilon^{ijk}~S^{jk}(x),~~~~~~~~~~L^i=\frac{1}{2}\int d^3x~\epsilon^{ijk}~L^{jk}(x),~~~~~~~~~~J^i=\frac{1}{2}\int d^3x~\epsilon^{ijk}~M^{jk}(x).\nonumber \\
\label{sot}
\eea
Using eqs. (\ref{ccnem}) and (\ref{hnem}) in (\ref{sot}) we find that the orbital angular momentum ${\vec L}_\gamma$ of the electromagnetic field is given by
\bea
{\bf L}_\gamma = \int d^3x~E^i(x) ~{\bf r} \times {\bf \nabla} A^i(x)
\label{dcnem}
\eea
which is not gauge invariant.

\subsection{ Gauge Non-Invariant Spin Angular Momentum of Electromagnetic Field in Maxwell Theory From Noether's Theorem }

Using eq. (\ref{bcnem}) in (\ref{sot}) we find that the spin angular momentum ${\vec S}_\gamma$ of the electromagnetic field is given by
\bea
{\bf S}_\gamma = \int d^3x ~{\bf E}(x)\times {\bf A}(x)
\label{ecnem}
\eea
which reproduces eq. (\ref{dspn}) which is not gauge invariant.

\subsection{ Gauge Non-Invariant Orbital Angular Momentum of Electron in Dirac Theory From Noether's Theorem }

The Dirac lagrangian density is given by
\bea
{\cal L} = \frac{1}{2}{\bar \psi}(x)[i\gamma^\lambda {\overrightarrow \partial}_\lambda -m -e\gamma^\lambda A_\lambda(x) ]\psi(x)-\frac{1}{2}{\bar \psi}(x)[i\gamma^\lambda {\overleftarrow \partial}_\lambda +m+e\gamma^\lambda A_\lambda(x)  ]\psi(x)
\label{kat}
\eea
which by using the Euler-Lagrange equation gives the Noether's theorem
\bea
\partial_\mu \frac{i}{2} [{\bar \psi}(x)\gamma^\mu \delta \psi(x)-[\delta {\bar \psi}(x)] \gamma^\mu \psi(x)]=0
\label{kgt}
\eea
where $\delta \psi(x)$ is the functional differential of the Dirac field $\psi(x)$ of the electron.

Under (infinitesimal) Lorentz transformation the Dirac spinors transform as
\bea
&& \psi'(x')=\psi(x)+\frac{1}{2}\epsilon_{\mu \nu} \frac{1}{2i}\sigma^{\mu \nu} \psi(x)\nonumber \\
&& {\bar \psi}'(x')={\bar \psi}(x)-\frac{1}{2}{\bar \psi}(x)\epsilon_{\mu \nu} \frac{1}{2i}\sigma^{\mu \nu}
\label{kht}
\eea
where
\bea
\sigma^{\mu \nu}=\frac{i}{2}[\gamma^\mu,~\gamma^\nu].
\label{kit}
\eea
Using eq. (\ref{kht}) in the functional differential of the Dirac spinor
\bea
\delta \psi(x) = \psi'(x') -\psi(x) -\delta x^\mu \partial_\mu \psi(x)
\label{cltrf}
\eea
we find
\bea
&&\delta \psi(x) =\frac{1}{2}\epsilon_{\mu \nu} \frac{1}{2i}\sigma^{\mu \nu} \psi(x)-\delta x^\mu \partial_\mu \psi(x) \nonumber \\
&& \delta {\bar \psi}(x) =-\frac{1}{2}{\bar \psi}(x)\epsilon_{\mu \nu}\frac{1}{2i} \sigma^{\mu \nu}-\delta x^\mu \partial_\mu {\bar \psi}(x).
\label{kjt}
\eea

Under translation [without rotation, $\epsilon_{\mu \nu}=0$, see eq. (\ref{gspn})] by using eq. (\ref{kjt}) in (\ref{kgt}) in Dirac theory we find the continuity equation
\bea
\partial_\mu T^{\mu \nu}(x)=0
\label{gnel}
\eea
where the gauge non-invariant energy-momentum tensor of the electron in Dirac theory obtained from Noether's theorem using Lorentz transformation is given by
\bea
&&T^{\mu \nu}(x)= \frac{i}{2} {\bar \psi}(x)[\gamma^\mu  {\overrightarrow \partial}^\nu -\gamma^\mu {\overleftarrow \partial}^\nu ] \psi(x).
\label{hnel}
\eea

Under rotation [without translation, $\Delta_\mu =0$, see eq. (\ref{gspn})] by using eq. (\ref{kjt}) in (\ref{kgt}) in Dirac theory we find the continuity equation
\bea
\partial_\mu J^{\mu \nu \lambda}(x)=0,~~~~~~~~~~~J^{\mu \nu \lambda}(x)=S^{\mu \nu \lambda}(x)- T^{\mu \nu}(x) x^\lambda+ T^{\mu \lambda}(x) x^\nu 
\label{bnd}
\eea
where
\bea
S^{\mu \nu \lambda}(x) = \frac{1}{4}{\bar \psi}(x) \{\gamma^\mu,~\sigma^{\nu \lambda} \} \psi(x).
\label{kpt}
\eea

From eq. (\ref{bnd}) the gauge non-invariant angular momentum tensor of the electron is given by
\bea
M^{\mu \nu}(x)=J^{0 \mu \nu}(x)=S^{\mu \nu }(x)+L^{\mu \nu }(x),~~~~~~~S^{\mu \nu}(x)=S^{0 \mu \nu}(x),~~~~~~L^{\mu \nu}(x)=- T^{0 \mu}(x) x^\nu + T^{0 \nu}(x) x^\mu \nonumber \\
\label{cnd}
\eea
where the gauge invariant $S^{\mu \nu \lambda}(x)$ is given by eq. (\ref{kpt}) and the gauge non-invariant $T^{\mu \nu}(x)$ is given by eq. (\ref{hnel}). By using eqs. (\ref{cnd}) and (\ref{hnel}) in (\ref{sot}) we find that the orbital angular momentum ${\vec L}_e$ of the electron is given by
\bea
{\bf L}_e = \int d^3x~{\bf r} \times [\psi^\dagger(x) [-i{\overrightarrow {\bf \partial}}+i{\overleftarrow {\bf \partial}}] \psi(x)]
\label{acnd}
\eea
which is not gauge invariant. 

Note that the gauge invariant definition of the orbital angular momentum of the electron in Dirac theory can be obtained from the gauge invariant Noether's theorem, see eq. (\ref{aonm}).

\subsection{ Gauge Invariant Spin Angular Momentum of Electron in Dirac Theory From Noether's Theorem }

By using eqs. (\ref{cnd}) and (\ref{kpt}) in (\ref{sot}) we find that the spin angular momentum ${\vec S}^{inv}_e$ of the electron is given by
\bea
{\bf S}^{inv}_e =  \int d^3x~\psi^\dagger(x) ~{\bf \Sigma}~ \psi(x)
\label{bcnd}
\eea
which is gauge invariant.

\section{ Non-Vanishing Boundary Term and Gauge Non-Invariant Spin and Orbital Angular Momentum in Dirac-Maxwell Theory }

The Lagrangian density in Dirac-Maxwell theory is given by
\bea
{\cal L} = \frac{1}{2}{\bar \psi}(x)[i\gamma^\lambda {\overrightarrow \partial}_\lambda -m -e\gamma^\lambda A_\lambda(x) ]\psi(x)-\frac{1}{2}{\bar \psi}(x)[i\gamma^\lambda {\overleftarrow \partial}_\lambda +m+e\gamma^\lambda A_\lambda(x)  ]\psi(x)-\frac{1}{4}F_{\mu \nu}(x)F^{\mu \nu}(x)\nonumber \\
\label{lelm}
\eea
and the Noether's theorem in Dirac-Maxwell theory is given by
\bea
\partial_\mu [\frac{i}{2} [{\bar \psi}(x)\gamma^\mu \delta \psi(x)-[\delta {\bar \psi}(x)] \gamma^\mu \psi(x)]+F^{\mu \nu}(x) \delta A_\nu(x)] - \delta x^\mu \partial_\mu {\cal L}(x)=0.
\label{nelm}
\eea
Under translation [without rotation, $\epsilon_{\mu \nu}=0$, see eq. (\ref{gspn})] we find by using
eqs. (\ref{fspn}) and (\ref{kjt}) in eq. (\ref{nelm}) in the Dirac-Maxwell theory the equation
\bea
\partial_\mu [ F^{\mu \lambda }(x) F_\lambda^{~~\nu}(x) -F^{\mu \lambda }(x)  \partial_\lambda A^\nu(x) +\frac{1}{4}g^{\mu \nu}  F_{\lambda \delta}(x)F^{\lambda \delta}(x)+\frac{i}{2} {\bar \psi}(x)[\gamma^\mu {\overrightarrow \partial}^\nu -\gamma^\mu {\overleftarrow \partial}^\nu  ] \psi(x)]=0\nonumber \\
\label{aiet}
\eea
which since $F^{\mu \lambda }(x)$ is antisymmetric in $\mu \leftrightarrow \lambda$ gives
\bea
\partial_\mu [ F^{\mu \lambda }(x) F_\lambda^{~~\nu}(x) -A^\nu(x) \partial_\lambda F^{\lambda \mu }(x)   +\frac{1}{4}g^{\mu \nu}  F_{\lambda \delta}(x)F^{\lambda \delta}(x)+\frac{i}{2} {\bar \psi}(x)[\gamma^\mu {\overrightarrow \partial}^\nu -\gamma^\mu {\overleftarrow \partial}^\nu  ] \psi(x)]=0.\nonumber \\
\label{celt}
\eea
Since
\bea
\partial_\lambda F^{\lambda \mu }(x)  =e{\bar \psi}(x) \gamma^\mu \psi(x)
\label{delt}
\eea
in Dirac-Maxwell theory we find from eq. (\ref{celt}) the continuity equation
\bea
\partial_\mu T^{\mu \nu}(x)=0
\label{biet}
\eea
where the gauge invariant energy-momentum tensor of the electromagnetic field plus electron in Dirac-Maxwell theory is given by
\bea
&&T^{\mu \nu}= F^{\mu \lambda }(x) F_\lambda^{~~\nu}(x) +\frac{1}{4}g^{\mu \nu}  F_{\lambda \delta}(x)F^{\lambda \delta}(x)+\frac{i}{2} {\bar \psi}(x)[\gamma^\mu  ({\overrightarrow \partial}^\nu +ieA^\nu(x)) -\gamma^\mu ({\overleftarrow \partial}^\nu -ieA^\nu(x))  ] \psi(x).\nonumber \\
\label{bkxt}
\eea
Note that even if the energy-momentum tensor $T^{\mu \nu}(x)$ in eq. (\ref{hnem}) of the electromagnetic field in Maxwell theory is not gauge invariant and
the energy-momentum tensor $T^{\mu \nu}(x)$ in eq. (\ref{hnel}) of the electron in Dirac theory is not gauge invariant but the energy-momentum tensor $T^{\mu \nu}(x)$ in eq. (\ref{bkxt}) of the electromagnetic field plus electron in Dirac-Maxwell theory is gauge invariant. This is because the gauge non-invariant part $\partial_\mu [F^{\mu \lambda }(x)  \partial_\lambda A^\nu(x)]$ from eq. (\ref{hnem}) when combined with the gauge non-invariant part  $\partial_\mu [\frac{i}{2} {\bar \psi}(x)[\gamma^\mu  {\overrightarrow \partial}^\nu -\gamma^\mu {\overleftarrow \partial}^\nu ] \psi(x)]$ from eq. (\ref{hnel}) gives the gauge invariant part $\partial_\mu [\frac{i}{2} {\bar \psi}(x)[\gamma^\mu  ({\overrightarrow \partial}^\nu +ieA^\nu(x)) -\gamma^\mu ({\overleftarrow \partial}^\nu -ieA^\nu(x))  ] \psi(x)]$ from eq. (\ref{bkxt}) [see eqs. (\ref{aiet}) -(\ref{bkxt}) for the derivation].

Under rotation [without translation, $\Delta_\mu =0$, see eq. (\ref{gspn})] we find by using eqs. (\ref{fspn}) and (\ref{kjt}) in eq. (\ref{nelm}) in the Dirac-Maxwell theory the equation
\bea
&&\partial_\mu [\frac{1}{2} S^{\mu \nu \lambda}(x)+x^\lambda [F^{\mu \delta }(x) F_\delta^{~~\nu}(x) -F^{\mu \delta }(x)  \partial_\delta A^\nu(x) +\frac{1}{4}g^{\mu \nu}  F_{\eta \delta}(x)F^{\eta \delta}(x)\nonumber \\
&&+\frac{i}{2} {\bar \psi}(x)[\gamma^\mu {\overrightarrow \partial}^\nu -\gamma^\mu {\overleftarrow \partial}^\nu  ] \psi(x)]-F^{\mu \lambda }(x)A^\nu(x)]\epsilon_{\nu \lambda} =0
\label{fiet}
\eea
which since $F^{\mu \delta }(x)$ is antisymmetric in $\mu \leftrightarrow \delta$ gives
\bea
&&\partial_\mu [\frac{1}{2}S^{\mu \nu \lambda}(x)+x^\lambda [F^{\mu \delta }(x) F_\delta^{~~\nu}(x) -A^\nu(x) \partial_\delta F^{\delta \mu }(x)  +\frac{1}{4}g^{\mu \nu}  F_{\eta \delta}(x)F^{\eta \delta}(x)\nonumber \\
&& +\frac{i}{2} {\bar \psi}(x)[\gamma^\mu {\overrightarrow \partial}^\nu -\gamma^\mu {\overleftarrow \partial}^\nu  ] \psi(x)]]\epsilon_{\nu \lambda} =0.
\label{giet}
\eea
Using eq. (\ref{delt}) in (\ref{giet}) we find the continuity equation
\bea
\partial_\mu J^{\mu \nu \lambda}(x)=0,~~~~~~~~~~J^{\mu \nu \lambda}(x)=S^{\mu \nu \lambda}(x)- T^{\mu \nu}(x) x^\lambda+ T^{\mu \lambda}(x) x^\nu.
\label{hielm}
\eea
From eq. (\ref{hielm}) the gauge invariant angular momentum tensor of the electron plus electromagnetic field in Dirac-Maxwell theory is given by
\bea
M^{\mu \nu}(x)=J^{0 \mu \nu}(x)
\label{jiet}
\eea
where the gauge invariant $S^{\mu \nu \lambda}(x)$ is given by eq. (\ref{kpt}) and the gauge invariant $T^{\mu \nu }(x)$ is given by eq. (\ref{bkxt}).

Using eqs. (\ref{jiet}), (\ref{hielm}), (\ref{bkxt}) and (\ref{kpt}) in (\ref{sot}) we find that the total angular momentum of the electron plus electromagnetic field in Dirac-Maxwell theory is given by
\bea
{\bf J}^{inv}_{\gamma +e} = \int d^3x~ [{\bf r} \times [{\bf E}(x) \times {\bf B}(x)] + \psi^\dagger(x) ~{\bf \Sigma}~ \psi(x)+ {\bf r} \times [\psi^\dagger(x) [-i{\overrightarrow {\bf D}}+i{\overleftarrow {\bf D}}] \psi(x)]]
\label{tanmge}
\eea
which is gauge invariant where the covariant derivative $D_\mu[A]$ is given by
\bea
D_\mu[A] =\partial_\mu + ie A_\mu(x).
\label{hspn}
\eea

\subsection{Non-Vanishing Boundary Term and Gauge Non-Invariant Spin and Orbital Angular Momentum of Electromagnetic Field and Gauge Non-Invariant Orbital Angular Momentum of Electron }

Note that using eq. (\ref{delt}) we find
\bea
&&  \int d^3x~[ {\bf r} \times [{\bf E}(x) \times {\bf B}(x)]+ {\bf r} \times [\psi^\dagger(x) [-i{\overrightarrow {\bf D}}+i{\overleftarrow {\bf D}}] \psi(x)]]\nonumber \\
&&=\int d^3x~ [{\bf E}(x)\times {\bf A}(x)+{ E}^l(x) {\bf r} \times {\bf \nabla} A^l(x) + {\bf r} \times [\psi^\dagger(x) [-i{\overrightarrow {\bf \partial}}+i{\overleftarrow {\bf \partial}}] \psi(x)]] \nonumber \\
&&- \int d^3x~ \partial_l [{ E}^l(x) {\bf r} \times {\bf A}(x)].
\label{ovbt}
\eea
In Maxwell theory the electromagnetic potential $A^\mu(x)$ produced at $x^\mu$ by the electron in motion at $X^\mu(\tau)$ with four-velocity $u^\mu(\tau)=\frac{dX^\mu(\tau)}{d\tau}$ is given by
\cite{nj1,ne1}
\bea
A^\nu(x) = e \frac{u^\nu(\tau_0)}{u(\tau_0) \cdot [x-X(\tau_0)]},~~~~~~~~~~~~~~~x_0-X_0(\tau_0)=|{\vec x}-{\vec X}(\tau_0)|
\label{lwp}
\eea
and the pure gauge potential $A^\mu_{pure}(x)$ is given by \cite{nj1,ne1}
\bea
A^\nu_{pure}(x) = e \frac{\beta_c^\nu}{\beta_c \cdot [x-X(\tau_0)]}=\partial^\nu \omega(x),~~~~~~~~~~~~~~~~\beta_c^2=0.
\label{pgp}
\eea
Hence in Dirac-Maxwell theory, because of the form in eq. (\ref{lwp}), we find the non-vanishing boundary term
\bea
\int d^3x~ \partial_l [{ E}^l(x) ~{\bf r} \times {\bf A}(x)] \neq 0
\label{pvbt}
\eea
which implies that for angular momentum the boundary term does not vanish in the Dirac-Maxwell theory although the similar boundary term vanishes for linear momentum.

Hence from eqs. (\ref{pvbt}) and (\ref{ovbt}) we find
\bea
&&  \int d^3x~[ {\bf r} \times [{\bf E}(x) \times {\bf B}(x)]+ {\bf r} \times [\psi^\dagger(x) [-i{\overrightarrow {\bf D}}+i{\overleftarrow {\bf D}}] \psi(x)]]\nonumber \\
&&\neq \int d^3x~ [{\bf E}(x)\times {\bf A}(x)+{ E}^l(x) {\bf r} \times {\bf \nabla} A^l(x) + {\bf r} \times [\psi^\dagger(x) [-i{\overrightarrow {\bf \partial}}+i{\overleftarrow {\bf \partial}}] \psi(x)]]
\label{qvbt}
\eea
see also \cite{bowdon}. Using eq. (\ref{qvbt}) in (\ref{tanmge}) we find
\bea
{\bf J}^{inv}_{\gamma + e} \neq {\bf S}_\gamma +{\bf L}_\gamma +{\bf S}^{inv}_e + {\bf L}_e
\label{totan}
\eea
where the gauge invariant total angular momentum ${\bf J}^{inv}_{\gamma +e}$ of the electron plus electromagnetic field in Dirac-Maxwell theory is given by eq. (\ref{tanmge}), the gauge non-invariant spin angular momentum ${\bf S}_\gamma$ of the electromagnetic field in Maxwell theory is given by eq. (\ref{ecnem}), the gauge non-invariant orbital angular momentum ${\bf L}_\gamma$ of the electromagnetic field in Maxwell theory is given by eq. (\ref{dcnem}), the gauge invariant spin angular momentum ${\bf S}^{inv}_e$ of the electron in Dirac theory is given by eq. (\ref{bcnd}) and the gauge non-invariant orbital angular momentum ${\bf L}_e$ of the electron in Dirac theory is given by eq. (\ref{acnd}).

\section{ Gauge Invariant Noether's Theorem in Maxwell Theory  }\label{gi}

In this section we derive the gauge invariant Noether's theorem in Maxwell theory by using the combined Lorentz transformation plus local gauge transformation. We obtain the gauge invariant and symmetric expression of the energy-momentum tensor from this gauge invariant Noether's theorem which predict the gauge invariant conserved energy-momentum of the electromagnetic field. Similarly, we obtain the gauge invariant expression of angular momentum tensor from this gauge invariant Noether's theorem which predict the gauge invariant conserved angular momentum of the electromagnetic field.

\subsection{ Combined Lorentz Transformation Plus Local Gauge Transformation in Maxwell Theory }

Under the local gauge transformation the $A^\mu(x)$ transforms as
\bea
A_\mu(x)  \rightarrow A^{GT}_\mu(x)=A_\mu(x) +\partial_\mu \omega(x)
\label{e5n}
\eea
which leaves the $F_{\mu \nu}(x)$ in eq. (\ref{p5n}) gauge invariant, {\it i. e.}.
\bea
F^{GT}_{\mu \nu}(x) =F_{\mu \nu}(x)
\label{hnt}
\eea
where $\omega(x)$ is any arbitrary function and the superscript symbol $GT$ means gauge transformed.

Under Lorentz transformation the four-vector $A_\mu(x)$ transforms as
\bea
A'_\mu(x') = \frac{\partial x^\nu}{\partial x'^\mu} A_\nu(x),~~~~~~~~~~~A'^\mu(x') = \frac{\partial x'^\mu}{\partial x^\nu} A^\nu(x)
\label{altr}
\eea
and the second rank tensor $F_{\mu \nu}(x)$ transforms as
\bea
F'^{\mu \nu}(x') =\frac{\partial x'^\mu}{\partial x^\lambda} \frac{\partial x'^\nu}{\partial x^\delta} F^{\lambda \delta}(x),~~~~~~~~F'_{\mu \nu}(x') =\frac{\partial x^\lambda}{\partial x'^\mu} \frac{\partial x^\delta}{\partial x'^\nu} F_{\lambda \delta}(x),~~~~~~~F'^\mu_{~~~\nu}(x') =\frac{\partial x'^\mu}{\partial x^\lambda} \frac{\partial x^\delta}{\partial x'^\nu} F^\lambda_{~~\delta}(x). \nonumber \\
\label{bltr}
\eea

From eq. (\ref{bltr}) we find that the $F_{\mu \nu}(x)$ has a well defined transformation of the tensor in Maxwell theory because the tensor $F_{\mu \nu}(x)$ is gauge invariant. However, this is not so for the vector $A_\mu(x)$ because this vector is not gauge invariant in Maxwell theory. From eq. (\ref{bltr}) we find that the general transformation of $A_\mu(x)$ is a Lorentz transformation plus gauge transformation given by
\bea
A'_\mu(x') =\frac{\partial x^\nu}{\partial x'^\mu} A_\nu(x) + \partial'_\mu \Lambda.
\label{v5nx}
\eea
Using eq. (\ref{v5nx}) in the functional differential
\bea
\delta A_\mu(x) = A'_\mu(x') -A_\mu(x) -\delta x^\nu \partial_\nu A_\mu(x)
\label{cltr}
\eea
we find
\bea
\delta A_\mu(x) = -\delta x^\nu F_{\nu \mu}(x)+ \partial_\mu [\Lambda-\delta x^\nu A_\nu(x)].
\label{htt}
\eea
Hence eq. (\ref{htt}) gives the functional differential $\delta A_\mu(x)$ when the combined Lorentz transformation plus local gauge transformation is used whereas eq. (\ref{fspn}) gives the functional differential $\delta A_\mu(x)$ when only the Lorentz transformation is used.

\subsection{ Derivation of Gauge Invariant Noether's Theorem in Maxwell Theory  }

Note that due to the presence of the additional parameter $\Lambda$ in the functional differential $\delta A_\mu(x)$ in eq. (\ref{htt}) it is possible to make the functional differential $\delta A_\mu(x)$ gauge invariant in eq. (\ref{htt}). However, since the local gauge transformation is not used in eq. (\ref{fspn}) we find that $\Lambda =0$ in eq. (\ref{fspn}) which implies that it is not possible to obtain gauge invariant functional differential $\delta A_\mu(x)$ when the Lorentz transformation is used alone. Hence one finds that it is not possible to derive the gauge invariant Noether's theorem from first principle when Lorentz transformation is used alone but it is possible to derive the gauge invariant Noether's theorem from first principle when combined Lorentz transformation plus local gauge transformation is used in gauge theory.

In order to derive the gauge invariant Noether's theorem in Maxwell theory from first principle by using combined Lorentz transformation plus local gauge transformation we proceed as follows. From eq. (\ref{htt}) we find
\bea
\delta A^{GT}_\mu(x) = -\delta x^\nu F^{GT}_{\nu \mu}(x)+ \partial_\mu [\Lambda^{GT}-\delta x^\nu A^{GT}_\nu(x)]
\label{hut}
\eea
which by using eqs. (\ref{e5n}) and (\ref{hnt}) gives
\bea
\delta A^{GT}_\mu(x) = -\delta x^\nu F_{\nu \mu}(x)+ \partial_\mu [\Lambda(A+\partial \omega)-\delta x^\nu A_\nu(x)-\delta x^\nu \partial_\nu \omega(x)].
\label{hvt}
\eea
For gauge invariant
\bea
\delta A_\mu(x)=\delta A^{GT}_\mu(x)
\label{hwt}
\eea
we find from eqs. (\ref{hvt}), (\ref{htt}) and (\ref{hwt}) that
\bea
\partial_\mu [\Lambda(A+\partial \omega)-\Lambda(A)-\delta x^\nu \partial_\nu \omega(x)]=0
\label{hxt}
\eea
which gives a simple solution
\bea
\Lambda = \delta x^\mu A_\mu(x)
\label{hyt}
\eea
which agrees with \cite{jackiw,eriksen,berg}. Using eq. (\ref{hyt}) in (\ref{htt}) we find
\bea
\delta A_\mu(x) = -\delta x^\nu F_{\nu \mu}(x)
\label{hzt}
\eea
which is gauge invariant.

Using eq. (\ref{hzt}) in (\ref{s5n1}) we find that the gauge invariant Noether's theorem in Maxwell theory obtained from the combined Lorentz transformation plus local gauge transformation is given by
\bea
\partial_\mu [-F^{\mu \nu}(x) F_{\lambda \nu}(x) \delta x^\lambda] -\delta x^\mu \partial_\mu {\cal L}(x)=0.
\label{iat}
\eea

\subsection{  Gauge Invariant Angular Momentum of Electromagnetic Field From Gauge Invariant Noether's Theorem in Maxwell Theory}

Under space-time translation (no rotation, $\epsilon_{\mu \nu}=0$) we find by using eqs. (\ref{gspn}) and (\ref{p5n}) in (\ref{iat}) the continuity equation
\bea
\partial_\mu T^{\mu \nu}(x)=0
\label{iet}
\eea
where the gauge invariant and symmetric energy-momentum tensor of the electromagnetic field in Maxwell theory is given by
\bea
T^{\mu \nu}(x)= F^{\mu \lambda }(x) F_\lambda^{~~\nu}(x)+\frac{1}{4}g^{\mu \nu}  F_{\lambda \delta}(x)F^{\lambda \delta}(x).
\label{igt}
\eea
Observe that we have not used the Belinfante tensor to make the energy-momentum tensor of the electromagnetic field in eq. (\ref{igt}) symmetric.

Hence we find that the gauge invariant and symmetric energy-momentum tensor of the electromagnetic field in eq. (\ref{igt}) in Maxwell theory is obtained from the first principle when the combined Lorentz transformation plus local gauge transformation is used to derive the Noether's theorem. The derivation of gauge invariant and symmetric energy-momentum tensor of electromagnetic field in eq. (\ref{igt}) by using combined Lorentz transformation plus local gauge transformation without using the Belinfante tensor implies that the Belinfante tensor is not required in Maxwell theory to make energy-momentum tensor of the electromagnetic field symmetric and gauge invariant.

Under rotation (no space-time translation, $\Delta_\mu=0$) we find by using eqs. (\ref{gspn}) and (\ref{p5n}) in (\ref{iat}) that
\bea
\partial_\mu [ T^{\mu \nu}(x) x^\lambda] \epsilon_{\nu \lambda}=0
\label{jmt}
\eea
where $T^{\mu \nu}(x)$ is given by eq. (\ref{igt}) which is gauge invariant and symmetric. Since
\bea
\epsilon_{\mu \nu}=-\epsilon_{\nu \mu}
\label{anst}
\eea
we find from eq. (\ref{jmt}) the continuity equation
\bea
\partial_\mu J^{\mu \nu \lambda}(x)=0
\label{jpt}
\eea
where the gauge invariant third rank tensor $J^{\mu \nu \lambda}(x)$ is given by
\bea
J^{\mu \nu \lambda}(x)=T^{\mu \lambda}(x) x^\nu - T^{\mu \nu}(x) x^\lambda
\label{jqt}
\eea
which gives the gauge invariant angular momentum tensor of the electromagnetic field in Maxwell theory
\bea
M^{\mu \nu }(x)=J^{0 \mu \nu}(x)=T^{0 \nu}(x) x^\mu - T^{0 \mu}(x) x^\nu
\label{kkqt}
\eea
where the gauge invariant and symmetric energy-momentum tensor $T^{\mu \nu}(x)$ of the electromagnetic field in Maxwell theory is given by eq. (\ref{igt}).

Hence we find that the gauge invariant angular momentum tensor of the electromagnetic field in eq. (\ref{kkqt}) in Maxwell theory is obtained from the first principle when the combined Lorentz transformation plus local gauge transformation is used to derive the Noether's theorem. The derivation of gauge invariant angular momentum tensor of the electromagnetic field in eq. (\ref{kkqt}) by using the combined Lorentz transformation plus local gauge transformation without using the Belinfante tensor implies that the Belinfante tensor is not required in Maxwell theory.

From eqs. (\ref{kkqt}), (\ref{igt}) and (\ref{sot}) we find that the gauge invariant definition of the angular momentum ${\vec J}^{inv}_e$of the electromagnetic field obtained from the gauge invariant Noether's theorem using combined Lorentz transformation plus local gauge transformation in Maxwell theory is given by
\bea
{\bf J}^{inv}_\gamma = \int d^3x~ {\bf r} \times [{\bf E}(x) \times {\bf B}(x)]
\label{atnm}
\eea
where ${\vec E}(x)$ is the electric field and ${\vec B}(x)$ is the magnetic field.

\section{ Gauge Invariant Noether's Theorem in Dirac Theory  }\label{gid}

In this section we derive the gauge invariant Noether's theorem in Dirac theory by using the combined Lorentz transformation plus local gauge transformation. We obtain the gauge invariant expression of the energy-momentum tensor from this gauge invariant Noether's theorem which predicts the gauge invariant conserved energy-momentum of the electron. Similarly, we obtain the gauge invariant expression of angular momentum tensor from this gauge invariant Noether's theorem which predicts the gauge invariant conserved angular momentum of the electron.

\subsection{ Combined Lorentz Transformation Plus Local Gauge Transformation in Dirac Theory }

Under (infinitesimal) local gauge transformation the Dirac field $\psi(x)$ of the electron transforms as
\bea
\psi(x)  \rightarrow \psi^{GT}(x)=\psi(x) -i e \omega(x) \psi(x)
\label{krt}
\eea
which along with the gauge transformation of $A_\mu(x)$ in eq. (\ref{e5n}) leaves the Dirac lagrangian density in eq. (\ref{kat})
gauge invariant.

From eq. (\ref{v5nx}) we find that under the combined Lorentz transformation plus gauge transformation the Dirac spinors transform as
\bea
&& \psi'(x')=\psi(x)+\frac{1}{2}\epsilon_{\mu \nu} \frac{1}{2i}\sigma^{\mu \nu} \psi(x)-ie \Lambda \psi(x) \nonumber \\
&& {\bar \psi}'(x')={\bar \psi}(x)-\frac{1}{2}{\bar \psi}(x)\epsilon_{\mu \nu} \frac{1}{2i}\sigma^{\mu \nu}+ {\bar \psi(x)} ie \Lambda
\label{kst}
\eea
where $\Lambda$ is given by eq. (\ref{hyt}). Using eqs. (\ref{kst}) and (\ref{hyt}) in (\ref{cltrf}) we find that under the combined Lorentz transformation plus local gauge transformation in Dirac theory the functional differentials of Dirac spinors are given by
\bea
&&\delta \psi(x) =\frac{1}{2}\epsilon_{\mu \nu} \frac{1}{2i}\sigma^{\mu \nu} \psi(x)-(\delta x^\mu) ({\overrightarrow \partial}_\mu +ieA_\mu(x)) \psi(x) \nonumber \\
&& \delta {\bar \psi}(x) =-\frac{1}{2}{\bar \psi}(x)\epsilon_{\mu \nu}\frac{1}{2i} \sigma^{\mu \nu}- {\bar \psi}(x) ({\overleftarrow \partial_\mu}-ieA_\mu(x))\delta x^\mu.
\label{ktt}
\eea
Hence we find that eq. (\ref{ktt}) gives the functional differential $\delta \psi(x)$ of the Dirac field $\psi(x)$ of the electron when the combined Lorentz transformation plus local gauge transformation is used whereas eq. (\ref{kjt}) gives the functional differential $\delta \psi(x)$ of the Dirac field $\psi(x)$ of the electron when only the Lorentz transformation is used.

\subsection{ Derivation of Gauge Invariant Noether's Theorem in Dirac Theory  }

Note that due to the presence of the covariant derivative as given by eq. (\ref{hspn})
in the functional differential $\delta \psi(x)$ in eq. (\ref{ktt}) it is possible to derive the gauge invariant Noether's theorem in Dirac theory when combined Lorentz transformation plus local gauge transformation is used.
However, since the local gauge transformation is not used in eq. (\ref{kjt}) we find that the functional differential $\delta \psi(x)$ contains the ordinary derivative $\partial_\mu$ (instead of the covariant derivative $D_\mu[A]$) in eq. (\ref{kjt}) which implies that it is not possible to obtain gauge invariant Noether's theorem when only the Lorentz transformation is used. Hence one finds that it is not possible to derive the gauge invariant Noether's theorem from first principle when Lorentz transformation is used alone but it is possible to derive the gauge invariant Noether's theorem from first principle when combined Lorentz transformation plus local gauge transformation is used.

In order to derive gauge invariant Noether's theorem in Dirac theory from first principle by using combined Lorentz transformation plus local gauge transformation we proceed as follows. Using eq. (\ref{ktt}) in (\ref{kgt}) we find that the gauge invariant Noether's theorem in Dirac theory is given by
\bea
\partial_\mu [\frac{1}{2} S^{\mu \nu \lambda}(x) \epsilon_{\nu \lambda} -\frac{i}{2}(\delta x_\nu) {\bar \psi}(x)\gamma^\mu [{\overrightarrow \partial}^\nu +ieA^\nu(x)] \psi(x)+
 \frac{i}{2}{\bar \psi}(x) [{\overleftarrow \partial^\nu}-ieA^\nu(x)] \gamma^\mu \psi(x) \delta x_\nu ]=0\nonumber \\
\label{kut}
\eea
where the gauge invariant third rank tensor $S^{\mu \nu \lambda}(x)$ is given by eq. (\ref{kpt}).

\subsection{  Gauge Invariant Orbital Angular Momentum of Electron From Gauge Invariant Noether's Theorem in Dirac Theory}

Under translation (without rotation, $\epsilon_{\mu \nu}=0$) we find by using eq. (\ref{gspn}) in gauge invariant Noether's theorem in Dirac theory in eq. (\ref{kut}) the continuity equation
\bea
\partial_\mu T^{\mu \nu}(x) = 0
\label{kwt}
\eea
where the gauge invariant energy-momentum tensor of the electron in Dirac theory is given by
\bea
T^{\mu \nu}(x)= \frac{i}{2} {\bar \psi}(x)[\gamma^\mu  ({\overrightarrow \partial}^\nu +ieA^\nu(x)) -\gamma^\mu ({\overleftarrow \partial}^\nu -ieA^\nu(x))  ] \psi(x).
\label{kxt}
\eea

Hence we find that the gauge invariant energy-momentum tensor in eq. (\ref{kxt}) of the electron in Dirac theory is obtained from the first principle when the combined Lorentz transformation plus local gauge transformation is used to derive the Noether's theorem.

Under rotation (without translation, $\Delta_\mu =0$) we find by using eqs. (\ref{gspn}) and (\ref{anst}) in the gauge invariant Noether's theorem in Dirac theory in eq. (\ref{kut}) the continuity equation
\bea
\partial_\mu J^{\mu \nu \lambda}(x)=0
\label{pat}
\eea
where the gauge invariant third rank tensor $J^{\mu \nu \lambda}(x)$ is given by
\bea
J^{\mu \nu \lambda}(x)=S^{\mu \nu \lambda}(x)- T^{\mu \nu}(x) x^\lambda+ T^{\mu \lambda}(x) x^\nu =S^{\mu \nu \lambda}(x)+L^{\mu \nu \lambda}(x)
\label{pbt}
\eea
which gives the gauge invariant angular momentum tensor of the electron
\bea
M^{\mu \nu}(x)=J^{0 \mu \nu}(x)=S^{0 \mu \nu}(x)- T^{0 \mu}(x) x^\nu + T^{0 \nu}(x) x^\mu
\label{pbtm}
\eea
where the gauge invariant $S^{\mu \nu \lambda}(x)$ is given by eq. (\ref{kpt}) and the gauge invariant $T^{\mu \nu}(x)$ is given by eq. (\ref{kxt}). From eq. (\ref{pbtm}) we find
\bea
M^{\mu \nu}(x) =S^{\mu \nu }(x)+L^{\mu \nu }(x)
\label{pbtm1}
\eea
where
\bea
S^{\mu \nu}(x)=S^{0 \mu \nu}(x),~~~~~~~~~~~~~~~~~L^{\mu \nu}(x)= T^{0 \nu}(x) x^\mu- T^{0 \mu}(x) x^\nu.
\label{pbtms}
\eea
Hence we find that the gauge invariant angular momentum tensor in eq. (\ref{pbtm}) of the electron in Dirac theory is obtained from the first principle when the combined Lorentz transformation plus local gauge transformation is used to derive the Noether's theorem.

From eqs. (\ref{pbtms}), (\ref{kxt}) and (\ref{sot}) we find that the gauge invariant definition of the orbital angular momentum ${\vec L}^{inv}_e$of the electron obtained from the gauge invariant Noether's theorem using combined Lorentz transformation plus local gauge transformation in Dirac theory is given by
\bea
{\bf L}^{inv}_e = \int d^3x~ {\bf r} \times [\psi^\dagger(x) [-i{\overrightarrow {\bf D}}+i{\overleftarrow {\bf D}}] \psi(x)]
\label{aonm}
\eea
where the covariant derivative $D_\mu[A]$ is given by eq. (\ref{hspn}).

\subsection{  Gauge Invariant Spin Angular Momentum of Electron From Gauge Invariant Noether's Theorem in Dirac Theory}

From eqs. (\ref{pbtms}), (\ref{kpt}) and (\ref{sot}) we find that the gauge invariant definition of the spin angular momentum ${\vec S}^{inv}_e$of the electron obtained from the gauge invariant Noether's theorem using combined Lorentz transformation plus local gauge transformation in Dirac theory is given by
\bea
{\bf S}^{inv}_e = \int d^3x~\psi^\dagger(x) ~{\bf \Sigma}~ \psi(x).
\label{asnm}
\eea

\section{ Gauge Invariant Noether's Theorem in Dirac-Maxwell Theory  }

We find by using eqs. (\ref{hzt}) and (\ref{ktt}) in eq. (\ref{nelm}) in the Dirac-Maxwell theory the gauge invariant Noether's theorem
\bea
&&\partial_\mu [\frac{1}{2} S^{\mu \nu \lambda}(x) \epsilon_{\nu \lambda} -\frac{i}{2}(\delta x_\nu) {\bar \psi}(x)\gamma^\mu [{\overrightarrow \partial}^\nu +ieA^\nu(x)] \psi(x)+ \frac{i}{2}{\bar \psi}(x) [{\overleftarrow \partial^\nu}-ieA^\nu(x)] \gamma^\mu \psi(x) \delta x_\nu\nonumber \\
&&-F^{\mu \nu}(x) F_{\lambda \nu}(x) \delta x^\lambda] -\delta x^\mu \partial_\mu {\cal L}(x)=0
\label{ntdm}
\eea
where the lagrangian density ${\cal L}(x)$ in the Dirac-Maxwell theory is given by eq. (\ref{lelm}).

Under translation [without rotation, $\epsilon_{\mu \nu}=0$] using eq. (\ref{gspn}) in (\ref{ntdm}) we find the continuity equation
\bea
\partial_\mu T^{\mu \nu}(x)=0
\label{tmdm}
\eea
where the gauge invariant energy-momentum tensor of the electromagnetic field plus  electron in Dirac-Maxwell theory obtained from the gauge invariant Noether's theorem using combined Lorentz transformation plus local gauge transformation is given by
\bea
&&T^{\mu \nu}(x)= F^{\mu \lambda }(x) F_\lambda^{~~\nu}(x) +\frac{1}{4}g^{\mu \nu}  F_{\lambda \delta}(x)F^{\lambda \delta}(x)+\frac{i}{2} {\bar \psi}(x)[\gamma^\mu  ({\overrightarrow \partial}^\nu +ieA^\nu(x)) -\gamma^\mu ({\overleftarrow \partial}^\nu -ieA^\nu(x))  ] \psi(x).\nonumber \\
\label{ckxt}
\eea

Under rotation [without translation, $\Delta_\mu =0$] we find by using eqs. (\ref{gspn}) in (\ref{ntdm}) the continuity equation
\bea
\partial_\mu J^{\mu \nu \lambda}(x)=0
\label{gieem}
\eea
where the gauge invariant third rank tensor $J^{\mu \nu \lambda}(x)$ is given by
\bea
J^{\mu \nu \lambda}(x)=S^{\mu \nu \lambda}(x)- T^{\mu \nu}(x) x^\lambda+ T^{\mu \lambda}(x) x^\nu=S^{\mu \nu \lambda}(x)+L^{\mu \nu \lambda}(x)
\label{hieem}
\eea
which gives the gauge invariant angular momentum tensor
\bea
M^{\mu \nu}(x)=J^{0 \mu \nu}(x)=S^{0 \mu \nu}(x)- T^{0 \mu}(x) x^\nu + T^{0 \nu}(x) x^\mu
\label{fieem}
\eea
of the electromagnetic field plus electron in Dirac-Maxwell theory obtained from the gauge invariant Noether's theorem using combined Lorentz transformation plus local gauge transformation where the gauge invariant $S^{\mu \nu \lambda}(x)$ is given by eq. (\ref{kpt}) and the gauge invariant $T^{\mu \nu}(x)$ is given by eq. (\ref{ckxt}).

\section{ Symmetric and Gauge Invariant Energy-Momentum Tensor in Dirac-Maxwell Theory Without Belinfante-Rosenfeld Tensor}

From eqs. (\ref{gieem}) and (\ref{hieem}) we find
\bea
\partial_\mu J^{\mu \nu \lambda}(x)=\partial_\mu [S^{\mu \nu \lambda}(x)+L^{\mu \nu \lambda}(x)]=0
\label{sbtm}
\eea
which implies
\bea
\partial_\mu L^{\mu \nu \lambda}(x)\neq 0
\label{tbtm}
\eea
where
\bea
L^{\mu \nu \lambda}(x)=T^{\mu \lambda}(x) x^\nu - T^{\mu \nu}(x) x^\lambda.
\label{lmu}
\eea
Hence from eqs. (\ref{tbtm}) and (\ref{lmu}) we find that the gauge invariant energy-momentum tensor $T^{\mu \nu}(x)$ of the electromagnetic field plus electron in eq. (\ref{ckxt}) in Dirac-Maxwell theory is not required to be symmetric although it is possible to obtain a symmetric and gauge invariant energy-momentum tensor $T_S^{\mu \nu}(x)$ in Dirac-Maxwell theory from the gauge invariant Noether's theorem without requiring Belinfante-Rosenfeld tensor.

In order to obtain a gauge invariant and symmetric energy-momentum tensor $T_S^{\mu \nu}(x)$ in Dirac-Maxwell theory from the gauge invariant Noether's theorem without requiring Belinfante-Rosenfeld tensor we proceed as follows. By using the Dirac lagrangian density from eq. (\ref{kat}) in the Euler-Lagrange equation we find the Dirac equation
\bea
&&i\gamma^\lambda {\overrightarrow \partial}_\lambda \psi(x) = [m +e\gamma^\lambda A_\lambda (x)]\psi(x), \nonumber \\
&& {\bar \psi}(x){\overleftarrow \partial}_\lambda  i\gamma^\lambda =-{\bar \psi}(x)[ m +e\gamma^\lambda A_\lambda (x)].
\label{ddrl}
\eea
From eqs. (\ref{ckxt}) and (\ref{ddrl}) we find by using the properties of the Dirac matrices the following equation
\bea
T^{\nu \mu}(x)-T^{\mu \nu}(x) =\partial_\lambda [\frac{1}{4}{\bar \psi}(x) \{\gamma^\lambda,~\sigma^{\mu \nu } \} \psi(x)].
\label{derl}
\eea
Using the properties of Dirac matrices we find
\bea
&&\{\gamma^\lambda, \sigma^{\mu \nu}\}=-\{\gamma^\lambda, \sigma^{\nu \mu}\},~~~~~~~~~\{\gamma^\lambda, \sigma^{\mu \nu}\}=-\{\gamma^\nu, \sigma^{\mu \lambda }\},~~~~~~~~~~\{\gamma^\lambda, \sigma^{\mu \nu}\}=-\{\gamma^\mu, \sigma^{\lambda \nu}\}\nonumber \\
\label{edrl}
\eea
which implies that the tensor ${\bar \psi}(x) \{\gamma^\lambda,~\sigma^{\mu \nu } \} \psi(x)$ is antisymmetric in $\lambda \leftrightarrow \mu$, $\lambda \leftrightarrow \nu$  and $\mu \leftrightarrow \nu$ which means
\bea
\partial_\mu \partial_\lambda [{\bar \psi}(x) \{\gamma^\lambda,~\sigma^{\mu \nu } \} \psi(x)]
=\partial_\nu \partial_\lambda [{\bar \psi}(x) \{\gamma^\lambda,~\sigma^{\mu \nu } \} \psi(x)]
=\partial_\mu \partial_\nu [{\bar \psi}(x) \{\gamma^\lambda,~\sigma^{\mu \nu } \} \psi(x)]=0.\nonumber \\
\label{dfrl}
\eea
We write eq. (\ref{tmdm}) as
\bea
\partial_\mu [\frac{1}{2}T^{\mu \nu}(x) +\frac{1}{2}T^{\nu \mu}(x)+\frac{1}{2}T^{\mu \nu}(x)-\frac{1}{2}T^{\nu \mu}(x)]= 0
\label{akwt}
\eea
which by using eqs. (\ref{derl}), (\ref{dfrl}) and (\ref{ckxt}) gives
\bea
\partial_\mu T_S^{\mu \nu}(x) =0
\label{akwtz}
\eea
where the symmetric and gauge invariant energy-momentum tensor $T_S^{\mu \nu}(x)$ of the electromagnetic field plus electron in Dirac-Maxwell theory is given by
\bea
&&T_S^{\mu \nu}(x)= F^{\mu \lambda }(x) F_\lambda^{~~\nu}(x) +\frac{1}{4}g^{\mu \nu}  F_{\lambda \delta}(x)F^{\lambda \delta}(x)+\frac{i}{4} {\bar \psi}(x)[\gamma^\mu  ({\overrightarrow \partial}^\nu +ieA^\nu(x)) +\gamma^\nu  ({\overrightarrow \partial}^\mu +ieA^\mu(x))\nonumber \\&&-\gamma^\mu ({\overleftarrow \partial}^\nu -ieA^\nu(x)) -\gamma^\nu ({\overleftarrow \partial}^\mu -ieA^\mu(x))  ] \psi(x)
\label{akxtz}
\eea
which gives
\bea
\partial_\mu J^{\mu \nu \lambda}(x)=\partial_\mu [T_S^{\mu \lambda}(x) x^\nu - T_S^{\mu \nu}(x) x^\lambda]=0.
\label{finem}
\eea

Hence we find that the gauge invariant and symmetric energy-momentum tensor in eq. (\ref{akxtz}) of the electromagnetic field plus electron in Dirac-Maxwell theory
is obtained from the first principle when the combined Lorentz transformation plus local gauge transformation is used to derive the Noether's theorem. The derivation of gauge invariant and symmetric energy-momentum tensor of the electromagnetic field plus electron in Dirac-Maxwell theory in (\ref{akxtz}) by using combined Lorentz transformation plus local gauge transformation without using the Belinfante-Rosenfeld tensor implies that the Belinfante-Rosenfeld tensor is not required in Dirac-Maxwell theory to make the energy-momentum tensor symmetric and gauge invariant.

\section{ Gauge Invariant Angular Momentum From Gauge Invariant Noether's Theorem in Dirac-Maxwell Theory  }

From eqs. (\ref{fieem}), (\ref{kpt}), (\ref{ckxt}) and (\ref{sot}) we find that the conserved and gauge invariant total angular momentum of the electromagnetic field plus electron obtained from the gauge invariant Noether's theorem in Dirac-Maxwell theory using combined Lorentz transformation plus local gauge transformation is given by
\bea
{\bf J}^{inv}_{\gamma +e} = \int d^3x~[ {\bf r} \times [{\bf E}(x) \times {\bf B}(x)] +\psi^\dagger(x) ~{\bf \Sigma}~ \psi(x) + {\bf r} \times [\psi^\dagger(x) [-i{\overrightarrow {\bf D}}+i{\overleftarrow {\bf D}}] \psi(x)]]
\label{tnm}
\eea
which agrees with eq. (\ref{tanmge}).

From eqs. (\ref{tnm}), (\ref{atnm}), (\ref{asnm}) and (\ref{aonm}) we find
\bea
{\bf J}^{inv}_{\gamma +e} = {\bf J}^{inv}_\gamma + {\bf S}^{inv}_e + {\bf L}^{inv}_e
\label{ttnmf}
\eea
which gives from eq. (\ref{totan})
\bea
{\bf J}^{inv}_{\gamma +e} = {\bf J}^{inv}_\gamma + {\bf S}^{inv}_e + {\bf L}^{inv}_e\neq {\bf S}_\gamma +{\bf L}_\gamma +{\bf S}^{inv}_e + {\bf L}_e
\label{ttnm}
\eea
where ${\bf J}^{inv}_{\gamma +e}$ is the gauge invariant conserved total angular momentum of the electromagnetic field plus electron, ${\bf J}^{inv}_{\gamma}$ is the gauge invariant total angular momentum of the electromagnetic field as given by eq. (\ref{atnm}), ${\bf S}^{inv}_{e}$ is the gauge invariant spin angular momentum of the electron as given by eq. (\ref{asnm}), ${\bf L}^{inv}_{ e}$ is the gauge invariant orbital angular momentum of the electron as given by eq. (\ref{aonm}), ${\bf S}_\gamma$ is the gauge non-invariant spin angular momentum of the electromagnetic field in Maxwell theory as given by eq. (\ref{ecnem}), ${\bf L}_\gamma$ is the gauge non-invariant orbital angular momentum of the electromagnetic field in Maxwell theory as given by eq. (\ref{dcnem}) and ${\bf L}_e$ is the gauge non-invariant orbital angular momentum of the electron in Dirac theory as given by eq. (\ref{acnd}).

Hence we find that since the difference between the angular momentum current built from the gauge invariant energy momentum tensor and that built in the naive Noether procedure is a total derivative, the space integral total angular momentum resulting from the two approaches differ by some surface term which may be finite for field configurations which do not decay fast enough at infinity.

From eq. (\ref{ttnm}) we find that the notion of the gauge invariant definition of the spin angular momentum ${\bf S}^{inv}_\gamma$ of the electromagnetic field (or the gauge invariant definition of the orbital angular momentum ${\bf L}^{inv}_\gamma$ of the electromagnetic field) does not exists in the Dirac-Maxwell theory.

Note that as mentioned earlier, as can be seen from eq. (\ref{ttnm}), the gauge invariant definition of the spin angular momentum of the electromagnetic field in the literature \cite{jaffe1,ji1,wata,hata,gold} is not correct because of the non-vanishing surface term [see eq. (\ref{pvbt})] in Dirac-Maxwell theory although the corresponding surface term vanishes for linear momentum.

\section{conclusions}

Due to proton spin crisis it is necessary to understand the gauge invariant definition of the spin and orbital angular momentum of the quark and gluon from first principle. In this paper we have derived the gauge invariant Noether's theorem by using combined Lorentz transformation plus local gauge transformation. We have found that the notion of the gauge invariant definition of the spin (or orbital) angular momentum of the electromagnetic field does not exist in Dirac-Maxwell theory although the notion of the gauge invariant definition of the spin (or orbital) angular momentum of the electron exists. We have found that the gauge invariant definition of the spin angular momentum of the electromagnetic field in the literature \cite{jaffe1,ji1,wata,hata,gold} is not correct because of the non-vanishing surface term [see eq. (\ref{pvbt})] in Dirac-Maxwell theory although the corresponding surface term vanishes for linear momentum. We have also shown that the Belinfante-Rosenfeld tensor is not required to obtain symmetric and gauge invariant energy-momentum tensor of the electron and the electromagnetic field in Dirac-Maxwell theory.

Note that the main idea of this paper is based on the expression (\ref{fspn}) of the variation of an abelian gauge field under an infinitesimal space time transformation $\delta A_\mu(x) = -F_{\nu \mu}(x)~ \delta x^\nu -\partial_\mu [A_\nu(x) ~\delta x^\nu]$. The second term is identified as a gauge variation so it can be dropped when applying the standard Noether procedure to the symmetry combined of the original space time transformation and this gauge transformation. This way one gets gauge invariant conserved currents like energy momentum tensor and angular momentum current. Similarly the same expression of $\delta A$ is true for the non-abelian gauge field with the derivative replaced by covariant derivative, so the same idea works also for the non-abelian case \cite{nkgym}. In particular, one finds for the non-abelian case $\delta A_\mu^b(x) =-\delta x^\nu \partial_\nu A_\mu^b(x)-A_\nu^b(x)\partial_\mu \delta x^\nu=-F_{\nu \mu}^b(x)~ \delta x^\nu -D_\mu[A_\nu^b(x) ~\delta x^\nu]$ where the non-abelian field tensor is given by $F_{\mu \nu}^b(x) =\partial_\mu A_\nu^b(x)-\partial_\nu A_\mu^b(x)+gf^{bda}A_\mu^d(x)A_\nu^a(x)$ and the covariant derivative is given by $D^{bd}_\mu=\delta^{bd} \partial_\mu +gf^{bad}A_\mu^a(x)$ with $a,b,d=1,...,8$ being the color indices \cite{nkgym}.

Hence we conclude that although the high energy collider experiments have measured the spin dependent gluon distribution function inside proton but we do not have a gauge invariant definition of the spin dependent gluon distribution function in QCD consistent with the gauge invariant Noether's theorem.

\end{document}